\documentclass[a4paper,11pt]{article}
\usepackage{latexsym}
\usepackage{amsmath}
\usepackage{amssymb}
\usepackage{amscd}

\newlength{\minitwocolumn}
\makeatletter
\@addtoreset{equation}{section}
\makeatother

\title{Form factors, correlation functions 
and vertex operators \\ 
in the eight-vertex model at reflectionless points\footnote{
Based on a talk given in the conference 
`Solvable Lattice Models 2004 
-- Recent Progress in Solvable Lattice Models --', 
RIMS, Kyoto University, 23 July 2004.}
}

\author{
Yas-Hiro Quano
\thanks{email: quanoy@suzuka-u.ac.jp}
}

\date{\it Department of Clinical Engineering, 
Suzuka University of Medical Science \\
\it Kishioka-cho 1001-1, Suzuka 510-0293, Japan}
\begin{document}

\maketitle

\begin{abstract}
The eight-vertex model at the reflectionless points is 
considered on the basis of Smirnov's axiomatic approach. 
Integral formulae for form factors of the eight-vertex model 
can be obtained in terms of those of the eight-vertex SOS model, 
by using vertex-face transformation. The resulting formulae 
have very simple forms at the reflectionless points, 
and suggest us the free field representation 
of type II vertex operators in the eight-vertex model. 
\end{abstract}

\section{Introduction}

In this paper we wish to construct 
form factors in the eight-vertex model 
at the reflectionless points. 
Form factors are originally defined as matrix elements 
of local operators. Through the study of form factors in the 
sine-Gordon model, Smirnov \cite{Smbk} 
found three axioms as sufficient conditions for the 
local commutativity of local fields in the model. 
Thus, following Smirnov, any objects 
that satisfy Smirnov's three axioms are referred 
to as `form factors'. 

For fixed a local operator ${\cal O}$, 
let 
\begin{eqnarray}
F^{(i)}_m ({\cal O}; \zeta_1, \cdots , \zeta_{2m})&=&
\displaystyle\sum_{\mu_1 \cdots \mu_{2m}} 
v^*_{\mu_1} \otimes \cdots \otimes 
v^*_{\mu_{2m}} 
F^{(i)}_m (\zeta_1 , \cdots , \zeta_{2m})_{
\mu_{1} \cdots \mu_{2m}}. \label{eq:v-ff}
\end{eqnarray}
Then Smirov's axioms are as follows \cite{Smbk}: 

1. $S$-matrix symmetry: 
\begin{eqnarray}
F^{(i)}_m (\cdots , \zeta_{j+1}, 
\zeta_{j}, \cdots )P_{j\,j+1}
&=&F^{(i)}_m (\cdots , \zeta_{j}, \zeta_{j+1}, \cdots )
S_{j\,j+1}(u_j -u_{j+1}), \label{eq:S-symm} 
\end{eqnarray}
where $\zeta_j =x^{-u_j}$, and $P$ is the permutation operator 
$(x\otimes y)P=y\otimes x$. 

2. cyclicity: 
\begin{equation}
F^{(i)}_m (\zeta', 
x^{-2}\zeta_{2m})=F^{(1-i)}_m (\zeta_{2m}, 
\zeta')P_{1\,2}\cdots 
P_{2m-1\,2m}. 
\label{eq:cyc-fi}
\end{equation}
where 
$\zeta' =(\zeta_1 , \cdots , \zeta_{2m-1})$.

3. annihilation pole condition
\begin{eqnarray}
&&
\underset{\zeta_{2m}=\varepsilon x^{-1}\zeta_{2m-1}}{\rm Res}\;
F^{(i)}_m (\zeta )
\dfrac{d\zeta_{2m}}{\zeta_{2m}} \label{eq:res-cond} \\
&=& \varepsilon^i \left( 
F^{(i)}_{m-1} (\zeta'' )
\otimes u^*_\varepsilon -F^{(1-i)}_{m-1} (\zeta'' )
\otimes u^*_\varepsilon 
S_{2m-1\;1}(u_{2m-1}-u_1 )\cdots 
S_{2m-1\;2m-2}(u_{2m-1}-u_{2m-2}) \right), \nonumber 
\end{eqnarray}
Here, $\zeta =(\zeta_1 , \cdots , \zeta_{2m})$, 
and 
$\zeta'' =(\zeta_1 , \cdots , \zeta_{2m-2})$; and $u^*_\varepsilon =
v^*_+ \otimes v^*_- +\varepsilon 
v^*_- \otimes v^*_+ $. 

The first two axioms imply the $q$-KZ equation \cite{FR} of level $0$: 
\begin{eqnarray}
F^{(i)}_m 
(\zeta_1,\cdots,x^{2}\zeta_j,\cdots,\zeta_{2m}) 
&=&
F^{(1-i)}_{m} (\zeta)
S_{j\,j+1}(u_{j}-u_{j+1})
\cdots
S_{j\,2m}(u_{j}-u_{2m}) \nonumber \\
&\times&
S_{j\,1}(u_{j}-u_1 -2) 
\cdots
S_{j\,j-1}(u_j-u_{j-1}-2). 
\label{eq:qKZ-0}
\end{eqnarray}

Lashkevich and Pugai \cite{LaP,LaP2} used the vertex-face 
correspondence \cite{3-bu/2} in order to 
construct the correlation functions 
of the eight-vertex/XYZ model in terms of those of 
the eight-vertex SOS model \cite{LuP}. The author constructed 
another simplified expression for the eight-vertex/XYZ 
correlation function, by solving Bootstrap equations \cite{qXYZ}. 
Shiraishi \cite{Shi} constructed 
the formulae of the correlation functions of the XYZ 
model without using the vertex-face correspondence. 

Concerning form factors in the eight-vertex model, 
Lashkevich \cite{La} found a bosonization recipe 
to construct integral representations of the form factors 
in the eight-vertex model. In principle, 
all form factors corresponding to all local fields 
can be constructed, but they take very complicated forms. 
We wish to construct simpler expressions in terms of 
the eight-vertex SOS model form factors \cite{SOSff}. 
This paper is the first trial for that purpose. 

\section{Basic definitions}

Let us consider the $Z$-invariant eight-vertex model \cite{Zinv} 
on a planar rectangular lattice. The state variables are 
associated with four edges around each vertex. Here 
the local state on an edge 
takes two possible values $(+)$ and $(-)$, respectively. 
The product of four states on the four edges around each vertex 
should be $+$ sign. This is called the generalized 
ice condition. 

Each straight line on the lattice carries 
a rapidity, or a spectral parameter. 
Let $V=\mathbb{C}v_+ \otimes \mathbb{C}v_-$, 
and let $V_u$ be a copy of $V$ with a rapidity $u$. 
Then the $R$-matrix $R^{V_{u_1}, V_{u_2}}$ 
can be regarded as an endomorphism on $V_{u_1}\otimes V_{u_2}$. 
It is due to the Lorentz invariance 
that $R^{V_{u_1}, V_{u_2}}$ depends only upon 
the difference of the rapidities $u_1 -u_2$. In what follows 
we thus denote $R^{V_{u_1}, V_{u_2}}$ by $R(u_1 -u_2 )$. 

The convention of the matrix elements of 
$R(u)\in \mbox{End}(V\otimes V)$ are as follows: 
\begin{equation}
R(u)v_{\varepsilon_1}\otimes v_{\varepsilon_2}=
\sum_{\varepsilon'_1,\varepsilon'_2 =\pm} 
v_{\varepsilon'_1}\otimes v_{\varepsilon'_2}
R(u)_{\varepsilon_1 \varepsilon_2}
    ^{\varepsilon'_1 \varepsilon'_2}. 
\label{eq:R-comp}
\end{equation}
For fixed $x=e^{-\epsilon}$ ($\epsilon >0$) and $r>1$, 
the explicit expression of the entries of $R(u)$ is 
given as follows:
\begin{equation}
R(u)=\frac{1}{\tilde{\kappa}(u)}\tilde{R}(u)=
\frac{1}{\bar{\kappa}(u)}\overline{R}(u)=\frac{1}{\bar{\kappa}(u)}
\begin{bmatrix} a(u) & & & d(u) \\
& b(u) & c(u) & \\
& c(u) & b(u) & \\ d(u) & & & a(u) \end{bmatrix}, 
\end{equation}
where 
\begin{eqnarray}
\tilde{\kappa} (u)&=& \dfrac{[1]}{[1-u]}\bar{\kappa}(u)
=\displaystyle\zeta^{\frac{r-1}{r}}
\frac{\rho (z)}{\rho (z^{-1})}, ~~~~ (z=\zeta^2 =x^{-2u}) 
\label{eq:kappa} \\
\rho (z)&=&\frac{(x^4 z ; x^4 , x^{2r})_{\infty}
      (x^{2r} z ; x^4 , x^{2r})_{\infty}
      }
     {(x^2 z ; x^4 , x^{2r})_{\infty}
            (x^{2r+2}z ; x^4 , x^{2r})_{\infty}}, ~~~~
(a;p_1,\cdots,p_n)_\infty=
\displaystyle \prod_{k_i\geqslant 0}
(1-ap_1^{k_1}\cdots p_n^{k_n}), \nonumber \\
{[} u {]} &=&\displaystyle x^{\frac{u^2}{r}-u}
\Theta_{x^{2r}}(x^{2u}), ~~~~ \Theta_p (z)=
(z; p)_\infty (pz^{-1}; p)_\infty (p; p)_\infty 
=\sum_{n\in\mathbb{Z}} p^{n(n-1)/2} (-z)^n, \nonumber 
\end{eqnarray}
\begin{equation}
\begin{array}{l}
a(u)=
\dfrac{\theta_2 
(\frac{u}{2r};\frac{\pi\sqrt{-1}}{2\epsilon r}) 
\theta_1 
(\frac{1-u}{2r};\frac{\pi\sqrt{-1}}{2\epsilon r})}
{\theta_2 (0;\frac{\pi\sqrt{-1}}{2\epsilon r}) 
\theta_1 
(\frac{1}{2r};
\frac{\pi\sqrt{-1}}{2\epsilon r})},
~~~~
b(u) 
=\dfrac{\theta_1(\frac{u}{2r};
\frac{\pi\sqrt{-1}}{2\epsilon r}) 
\theta_2 (\frac{1-u}{2r};
\frac{\pi\sqrt{-1}}{2\epsilon r})}
{\theta_2 (0;\frac{\pi\sqrt{-1}}{2\epsilon r}) 
\theta_1 
(\frac{1}{2r};
\frac{\pi\sqrt{-1}}{2\epsilon r})},
\\[6mm]
c(u)
=\dfrac{\theta_2 
(\frac{u}{2r};\frac{\pi\sqrt{-1}}{2\epsilon r}) 
\theta_2 
(\frac{1-u}{2r};\frac{\pi\sqrt{-1}}{2\epsilon r})}
{\theta_2 (0;\frac{\pi\sqrt{-1}}{2\epsilon r}) 
\theta_2 
(\frac{1}{2r};
\frac{\pi\sqrt{-1}}{2\epsilon r})},
~~~~
d(u)
=-\dfrac{\theta_1 
(\frac{u}{2r};\frac{\pi\sqrt{-1}}{2\epsilon r}) 
\theta_1 
(\frac{1-u}{2r};\frac{\pi\sqrt{-1}}{2\epsilon r})}
{\theta_2 (0;\frac{\pi\sqrt{-1}}{2\epsilon r}) 
\theta_2 
(\frac{1}{2r};
\frac{\pi\sqrt{-1}}{2\epsilon r})}, 
\end{array}
\label{eq:abcd}
\end{equation}
$$
\theta_{1,2} (u;\tau )=\sqrt{\mp 1}q^{\frac{1}{4}}
\zeta^{-1}
\Theta_{q^2} (\pm z), ~~~~ 
\theta_{3,4} (u;\tau )=
\Theta_{q^2} (\mp qz). ~~~~ (q=e^{\sqrt{-1}\pi\tau}, ~
z=\zeta^2 =e^{2\sqrt{-1}\pi u}) 
$$

The most important property of the $R$-matrix is the 
Yang-Baxter equation \cite{ESM}: 
\begin{equation}
R_{12}(\zeta_1/\zeta_2)
R_{13}(\zeta_1/\zeta_3)
R_{23}(\zeta_2/\zeta_3)=
R_{23}(\zeta_2/\zeta_3)
R_{13}(\zeta_1/\zeta_3)
R_{12}(\zeta_1/\zeta_2), 
\label{eq:YBE}
\end{equation}
where the subscript of the $R$-matrix denotes 
the spaces on which $R$ nontrivially acts. 

{}From (\ref{eq:kappa}) 
$\tilde{\kappa} (0)=1=\bar{\kappa} (0)$. It is easy to see 
$\bar{\kappa}(1-u)=\bar{\kappa}(u)$, which implies 
$\bar{\kappa}(1)=1$. From these and (\ref{eq:abcd}) we have 
\begin{equation}
R(0)=P=\begin{bmatrix} 1 & & & \\
& & 1 & \\
& 1 & & \\
& & & 1 \end{bmatrix}, ~~~~ R(1)=\begin{bmatrix} & & & \\
& 1 & 1 & \\
& 1 & 1 & \\
& & & \end{bmatrix}. 
\label{eq:R-sp}
\end{equation}
The unitarity relation
\begin{equation}
R_{12}(u)R_{21}(-u)=1, 
\label{eq:uni}
\end{equation}
and the crossing symmetries
\begin{equation}
R_{21}^{t_1}(1-u)=\sigma^x_1 
    R_{12}(u)\sigma^x_1 
\label{eq:cross}
\end{equation}
are also important. 

For fixed $\epsilon >0$ and $r>1$, the region $0<u<1$ 
is called the principal regime. This regime is one of 
antiferroelectric regions because of $c>a+b+|d|$. 
In the low temperature limit $\epsilon \rightarrow +\infty$ 
($c\gg a+b+|d|$), only $c$-type configuration is permitted 
at each vertex. Thus, there are two ground states in 
the principal regime. 

Let us consider the half infinite 
pure tensor vector $\cdots \otimes v_{\varepsilon_3} 
\otimes v_{\varepsilon_2} \otimes 
v_{\varepsilon_1}$ along a half infinite row. 
The ground state corresponds to the sequence 
$\varepsilon_j =(-1)^{j+i}$ ($i=0,1$). Fix the 
ground state labeled by $i$. Then 
at finite temperature $\epsilon >0$, 
any state configurations differ from that of $i$-th 
ground states by altering a finite number of spins. 
Otherwise, the system has infinitely high energy. 
Thus, the space of states ${\cal H}_i$ is the 
subspace of `$\cdots \otimes V \otimes V 
\otimes V$' spanned by 
$$
\cdots \otimes v_{\varepsilon_3} 
\otimes v_{\varepsilon_2} \otimes 
v_{\varepsilon_1}, ~~~~ 
\varepsilon_j =(-1)^{j+i} \;\; (j\gg 1). 
$$ 

Let us remind 
the definitions of the eight-vertex SOS model and 
the intertwining vectors. 
The eight-vertex SOS model is a face model \cite{3-bu/2} which is 
defined on the square lattice with a site variable 
$k_j \in \mathbb{Z}$ attached to each site $j$. 
We call $k_j$ a local state or a height 
and impose the condition 
that heights of adjoining sites differ by one. 
Local Boltzmann weight of this model is given for a state configuration 
$\displaystyle \begin{array}{c} 
\raisebox{1mm}{$c$} \\ \raisebox{-2mm}{$b$} \end{array} 
\fbox{\rule[-3mm]{0cm}{6mm}~~~~~} \!\begin{array}{c} 
\raisebox{1mm}{$d$} \\ \raisebox{-2mm}{$a$} \end{array}$ 
round a face. 
Here the four states $a, b, c$ and $d$ are ordered clockwise 
from the SE corner. ~~ 
The weights are assumed 
to be the functions of the spectral parameter $u$ and 
the nonzero Boltzmann weights are given as follows: 
\begin{equation}
\begin{array}{rcl}
\displaystyle W\left[ \left. \begin{array}{cc} 
k\pm 2 & k\pm 1 \\ k\pm 1 & k \end{array} \right| u \right]
& = & \dfrac{1}{\tilde{\kappa }(u)}=\dfrac{1}{\bar{\kappa }(u)}
\dfrac{[1-u]}{[1]}, \\[6mm]
\displaystyle W\left[ \left. \begin{array}{cc} 
k & k\pm 1 \\ k\pm 1 & k \end{array} \right| u \right]
& = & \displaystyle \dfrac{1}{\tilde{\kappa }(u)}
\frac{[1][k \pm u]}{[1-u][k]}=\dfrac{1}{\bar{\kappa }(u)}
\frac{[k \pm u]}{[k]}, \\[6mm]
\displaystyle W\left[ \left. \begin{array}{cc} 
k & k\mp 1 \\ k\pm 1 & k \end{array} \right| u \right]
& = & \displaystyle \dfrac{1}{\tilde{\kappa }(u)}
\frac{[u][k\pm 1]}{[1-u][k]}=\dfrac{1}{\bar{\kappa }(u)}
\frac{[u][k\pm 1]}{[1][k]}. 
\end{array}
\label{eq:HY}
\end{equation}

In regime III ($0<u<1$) the ground state of the eight-vertex 
SOS model (\ref{eq:HY}) can be labeled by an integer $l$, 
whose local states are $l$ or $l+1$. 
In what follows, we fix one of the ground state (labeled by, 
say, $l$), and consider any configurations which differ from 
that of the $l$-th ground state 
by changing a finite number of local states. 
Let us call a path $p=(k_1 , k_2 , k_3 , \cdots )$ an admissible 
path, if $|k_{j+1}-k_j|=1$ ($j=1,2,3,\cdots$) holds. Let 
${\cal H}_{l,k}^{(i)}$ ($i=0,1$) be the space of admissible 
paths satisfying the initial condition $k_1 =k$ and 
the following boundary condition 
$$
k_j =\left\{ \begin{array}{lll} 
l & \mbox {if $j\equiv 1-i$} & \mbox{(mod $2$)} \\
l+1 & \mbox {if $j\equiv i$} & \mbox{(mod $2$)} 
\end{array} \right. ~~~~ (j\gg 1). 
$$
Note that $i\equiv k-l$ (mod $2$).

The intertwining vectors 
\begin{equation}
t_{k}^{k\pm 1} (u)^{\varepsilon} 
=\frac{(\sqrt{-1})^{k-l+1/2}\varepsilon^{k-l}}{\sqrt{2}} 
f(u) \theta_{\bar{\varepsilon}} 
\left( 
\tfrac{k\mp u}{2r};\tfrac{\pi\sqrt{-1}}{2\epsilon r}\right), ~~~~
(\overline{+},\overline{-})=(3,4), 
\label{eq:int-vec-AP}
\end{equation}
map the eight-vertex SOS model in regime III onto the eight-vertex model 
in principal regime. Here, 
the normalization factor $f(u)$ satisfies the relation 
\begin{equation}
[u] f(u)f(u-1)=\frac{\pi}{\epsilon r} e^{\frac{\epsilon r}{2}}. 
\label{eq:C-df}
\end{equation}
The explicit expression of $f(u)$ is as follows: 
\begin{equation}
f(u)=\frac{x^{-\frac{u^2}{2r}+\frac{r-1}{2r}u+\frac{1}{4}}}
{C \sqrt{(x^{2r}; x^{2r})_\infty}} 
\frac{(x^{4+2u}; x^4, x^{2r})_\infty 
(x^{2r+2-2u}; x^4, x^{2r})_\infty}
{(x^{2+2u}; x^4, x^{2r})_\infty (x^{2r-2u}; x^4, x^{2r})_\infty}
\label{eq:f-def}
\end{equation}

Then we have the so-called vertex-face correspondence: 
\begin{equation}
\displaystyle R(u_1 -u_2 ) 
t^{c}_{b}(u_1 )\otimes t^{b}_{a}(u_2 ) 
=\displaystyle 
\sum_{d} W\left[ \left. \begin{array}{cc} 
c & d \\ b & a \end{array} \right| u_1 -u_2 \right] 
t^{d}_{a}(u_1 )\otimes t^{c}_{d}(u_2 ). 
\label{eq:JMO-AP}
\end{equation}

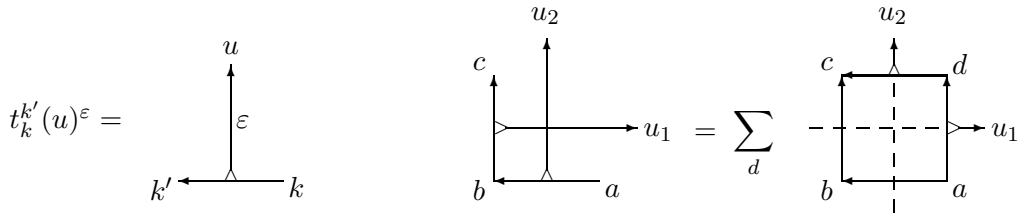
\unitlength 1.4mm
\begin{picture}(100,20)
\put(20,3){\begin{picture}(101,0)
\put(-16,5){$t^{k'}_{k}(u)^\varepsilon =$} 
\put(10.5,-1.5){$k$}
\put(-2.7,-1.8){$k'$}
\put(4.3145,0.){\scriptsize{$\wedge$}}
\put(5.5,5){$\varepsilon$}
\put(10,0){\vector(-1,0){10}}
\put(5,1){\vector(0,1){10}}
\put(4.2,12.){$u$}
\end{picture}
}
\put(50,3){\begin{picture}(101,0)
\put(10,0){\vector(-1,0){10}}
\put(0,0){\vector(0,1){10}}
\put(1.1,5){\vector(1,0){12.5}}
\put(-0.1,4.55){\scriptsize{$>$}}
\put(14.2,4.2){$u_1$}
\put(4.3145,0.){\scriptsize{$\wedge$}}
\put(5,1){\vector(0,1){12.5}}
\put(3.5,15.3){$u_2$}
\put(-2,10.5){$c$}
\put(10.5,-1.5){$a$}
\put(-2.,-1.8){$b$}
\put(19,4){$=\;\displaystyle\sum_{d}$} 
\end{picture}
}
\put(83,3){\begin{picture}(101,0)
\put(10,0){\vector(-1,0){10}}
\put(10,0){\vector(0,1){10}}
\put(0,0){\vector(0,1){10}}
\put(10,10){\vector(-1,0){10}}
\multiput(-3,5)(2.2,0){6}{\line(1,0){1.2}}
\put(9.9,4.55){\scriptsize{$>$}}
\put(11.1,5){\vector(1,0){2.5}}
\put(14.2,4.2){$u_1$}
\multiput(5,-3)(0,2.2){6}{\line(0,1){1.2}}
\put(4.3145,10.){\scriptsize{$\wedge$}}
\put(5,11){\vector(0,1){2.5}}
\put(3.5,15.3){$u_2$}
\put(10.5,10.1){$d$}
\put(-2.,10.5){$c$}
\put(10.5,-1.5){$a$}
\put(-2.,-1.8){$b$}
\end{picture}
}
\end{picture}

Let us introduce the following dual intertwining vectors: 
\begin{equation}
\begin{array}{l}
t^{*k'}_{k} (u; \epsilon , r)=t^{*k'}_{k} (u)
=\displaystyle\sum_{\varepsilon =\pm} 
t^{*k'}_{k} (u)_\varepsilon 
v^*_\varepsilon , \\ 
t^{*k'}_{k} (u)_\varepsilon 
=\dfrac{1}{[k]} t^{k'}_{k} (u+1)^{-\varepsilon}. 
\end{array} 
\end{equation}
{}From the following inversion relations 
\begin{equation}
\sum_{\varepsilon =\pm} t^{*k'}_{k} (u)_\varepsilon 
t^{k}_{k''} (u)^\varepsilon =\delta_{k''}^{k'}, ~~~~ 
\sum_{k'=k\pm 1} t^{k}_{k'} (u)^\varepsilon 
t^{*k'}_{k} (u)_{\varepsilon'} =
\delta^\varepsilon_{\varepsilon'}, \label{eq:dual-t}
\end{equation}
the dual vertex-face correspondence holds: 
\begin{equation}
t_{d}^{*a}(u_1 )\otimes t_{c}^{*d}(u_2 )
R(u_1 -u_2 ) =
\displaystyle\sum_{b} 
W\left[ \left. \begin{array}{cc} 
c & d \\ b & a \end{array} \right| u_1 -u_2 \right]
t_{c}^{*b}(u_1 )\otimes t_{b}^{*a}(u_2 ). 
\label{eq:dJMO}
\end{equation}

\unitlength 1.4mm
\begin{picture}(100,20)
\put(20,3){\begin{picture}(101,0)
\put(-16,5){$t^{*k'}_{k}(u)_\varepsilon =$} 
\put(10.5,10.5){$k'$}
\put(-2.7,10.8){$k$}
\put(10,10){\vector(-1,0){10}}
\put(5.5,5){$\varepsilon$}
\put(4.3145,8.8){\scriptsize{$\vee$}}
\put(5,0){\line(0,1){8.8}}
\put(5,10){\line(0,1){1}}
\put(5,12){\vector(0,1){1.5}}
\put(4.2,14.5){$u$}
\end{picture}
}
\put(50,3){\begin{picture}(101,0)
\put(10,0){\vector(0,1){10}}
\put(10,10){\vector(-1,0){10}}
\put(0,5){\line(1,0){8.8}}
\put(10,5){\line(1,0){1}}
\put(12,5){\vector(1,0){1.5}}
\put(8.5,4.55){\scriptsize{$<$}}
\put(14.2,4.2){$u_1$}
\put(4.3145,8.8){\scriptsize{$\vee$}}
\put(5,0){\line(0,1){8.8}}
\put(5,10){\line(0,1){1}}
\put(5,12){\vector(0,1){1.5}}
\put(3.5,15.3){$u_2$}
\put(-2,10.5){$c$}
\put(10.5,-1.5){$a$}
\put(10.5,10.1){$d$}
\put(19,4){$=\;\displaystyle\sum_{d}$} 
\end{picture}
}
\put(83,3){\begin{picture}(101,0)
\put(10,0){\vector(-1,0){10}}
\put(10,0){\vector(0,1){10}}
\put(0,0){\vector(0,1){10}}
\put(10,10){\vector(-1,0){10}}
\multiput(0,5)(2.2,0){5}{\line(1,0){1.2}}
\put(-3,5){\line(1,0){1.7}}
\put(-1.5,4.55){\scriptsize{$<$}}
\put(11.1,5){\vector(1,0){2.5}}
\put(14.2,4.2){$u_1$}
\multiput(5,0)(0,2.2){5}{\line(0,1){1.2}}
\put(5,-3){\line(0,1){1.8}}
\put(4.3145,-1.2){\scriptsize{$\vee$}}
\put(5,11){\vector(0,1){2.5}}
\put(3.5,15.3){$u_2$}
\put(10.5,10.1){$d$}
\put(-2.,10.5){$c$}
\put(10.5,-1.5){$a$}
\put(-2.,-1.8){$b$}
\end{picture}
}
\end{picture}

We also introduce another dual intertwining vector 
\begin{equation}
\begin{array}{l}
\displaystyle\tilde{t}^{*k'}_{k} (u; \epsilon , r)=
\tilde{t}^{*k'}_{k} (u)
=\displaystyle\sum_{\varepsilon =\pm} 
\tilde{t}^{*k'}_{k} (u)_\varepsilon 
v^*_\varepsilon , \\[2mm]
\displaystyle \tilde{t}^{*k'}_{k} (u)_\varepsilon 
=\dfrac{1}{[k']} 
t^{k}_{k'} (u-1)^{-\varepsilon}, 
\end{array}
\end{equation}
that satisfies the following inversion relations: 
\begin{equation}
\sum_{\varepsilon =\pm} \tilde{t}_{k'}^{*k} (u)_\varepsilon 
t^{k''}_{k} (u)^\varepsilon =\delta^{k''}_{k'}, ~~~~ 
\sum_{k'=k\pm 1} t_{k}^{k'} (u)^\varepsilon 
\tilde{t}_{k'}^{*k} (u)_{\varepsilon'} =
\delta^\varepsilon_{\varepsilon'}. \label{eq:dual-t'}
\end{equation}

\unitlength 1.4mm
\begin{picture}(100,20)
\put(10,3){\begin{picture}(101,0)
\put(10.5,-1.5){$k''$}
\put(-2.2,-1.8){$k$}
\put(4.3145,0.1){\scriptsize{$\wedge$}}
\put(10,0){\vector(-1,0){10}}
\multiput(0,0)(0,2.2){5}{\line(0,1){1.2}}
\put(10.5,10.5){$k'$}
\put(-2.2,10.4){$k$}
\put(10,10){\vector(-1,0){10}}
\put(4.3145,8.8){\scriptsize{$\vee$}}
\put(5,1.2){\line(0,1){7.6}}
\put(15.,4.5){$=\delta_{k''}^{k'}=$}
\end{picture}
}
\put(40,3){\begin{picture}(101,0)
\put(10.5,-1.5){$k$}
\put(-3.2,-1.8){$k''$}
\put(4.3145,0.1){\scriptsize{$\wedge$}}
\put(10,0){\vector(-1,0){10}}
\multiput(10,0)(0,2.2){5}{\line(0,1){1.2}}
\put(10.5,10.5){$k$}
\put(-2.2,10.4){$k'$}
\put(10,10){\vector(-1,0){10}}
\put(5,10){\oval(1.2,1.2)[b]}
\put(5,1.2){\line(0,1){8.2}}
\end{picture}
}
\put(85,3){\begin{picture}(101,0)
\put(-16,5){$\tilde{t}^{*k'}_{k}(u)_\varepsilon =$} 
\put(10.5,10.5){$k'$}
\put(-2.7,10.8){$k$}
\put(10,10){\vector(-1,0){10}}
\put(10,10){\vector(-1,0){10}}
\put(5,10){\oval(1.2,1.2)[b]}
\put(5.5,5){$\varepsilon$}
\put(5,0){\line(0,1){9.3}}
\put(5,10){\line(0,1){1}}
\put(5,12){\vector(0,1){1.5}}
\put(4.2,14.5){$u$}
\end{picture}
}
\end{picture}

For fixed $r>1$, let 
\begin{equation}
S(u)=-R(u; \epsilon , r-1), ~~~~
W'\left[ \left. 
\begin{array}{cc} c & d \\ b & a \end{array} \right| 
u \right]=-W\left[ \left. 
\begin{array}{cc} c & d \\ b & a \end{array} \right| 
u \right] \left. \makebox{\rule[-4mm]{0pt}{11mm}} 
\right|_{r\mapsto r-1}, 
\label{eq:SXYZ}
\end{equation}
and 
\begin{equation}
t'{}^{*k}_{k'} (u):=t^{*k}_{k'} (u; \epsilon , r-1). 
\label{eq:t'*}
\end{equation}
Then we have 
\begin{equation}
\displaystyle t'{}_{d}^{*a}(u_1 )\otimes 
t'{}_{c}^{*d}(u_2 )S(u_1 -u_2 ) 
=\displaystyle\sum_{d} 
W'\left[ \left. \begin{array}{cc} 
c & d \\ b & a \end{array} \right| u_1 -u_2 \right]
t'{}_{c}^{*b}(u_1 )\otimes t'{}_{b}^{*a}(u_2 ). 
\label{eq:sJMO}
\end{equation}
Note that the normalization factor in (\ref{eq:SXYZ}) is given by 
\begin{equation}
\tilde{\kappa}^* (\zeta )=-\tilde{\kappa} (\zeta ; x, x^{2(r-1)})
=\zeta^{-\frac{r}{r-1}} 
\frac{g^*(z)}{g^*(z^{-1})}, ~~~~ 
g^* (z)=g^* (x^4 z^{-1}). 
\label{eq:g*-prop}
\end{equation}
The explicit expression of $g^* (z)$ is given as follows: 
\begin{equation}
g^* (z)=\frac{
\{ z\}'_\infty \{ x^4 z^{-1}\}'_\infty 
\{ x^{2r+2}z\}'_\infty \{ x^{2r+6}z^{-1}\}'_\infty}
{\{ x^2 z\}'_\infty \{ x^6 z^{-1}\}'_\infty 
\{ x^{2r}z\}'_\infty \{ x^{2r+4}z^{-1}\}'_\infty}, ~~~~
\; \{ z\}'_\infty =(z; x^4 , x^4 , x^{2(r-1)}). 
\label{eq:g*}
\end{equation}

The eight-vertex model is on the `reflectionless point' 
if $r=1+1/N$ ($N=1,2,3,\cdots$) and therefore 
the $S$-matrix becomes (anti-)diagonal. When $r=2$ ($N=1$) 
the XYZ model is equivalent to the double Ising model 
\cite{ESM}, as is well known. 

\section{Form factors in the eight-vertex SOS model}

In this section 
we construct integral formulae for 
form factors in the eight-vertex SOS model. 
The first two axioms for 
form factors in the eight-vertex SOS model are as follows: 

1. $W'$--symmetry
\begin{equation}
\begin{array}{cl}
&F^{(l,k)}_{m} 
(\cdots,\zeta_{j+1},\zeta_j,\cdots)_{\cdots 
l_{j-1}l_jl_{j+1} \cdots} \\[4mm]
=&\displaystyle\sum_{l'_j} 
W'\left[ \left. \begin{array}{cc} 
l_{j+1} & l'_j \\ l_j & l_{j-1} \end{array} \right| 
u_j-u_{j+1} \right] F^{(l,k)}_{m} 
(\cdots,\zeta_j,\zeta_{j+1},\cdots)_{\cdots 
l_{j-1}l'_jl_{j+1} \cdots}. 
\end{array}
\label{eq:W'-symm-comp} 
\end{equation}

2. Cyclicity
\begin{equation}
F^{(l,k)}_{m} 
(\zeta', x^{-2}\zeta_{2m})_{
l\cdots l'l}
=F^{(l',k)}_{m} 
(\zeta_{2m}, \zeta')_{
l'l\cdots l'}. 
\label{eq:F'cyc-comp}
\end{equation}
Here, we only consider the case $l_0 =l_{2m}$ for 
$2m$-pt SOS form factors. These two imply the 
$q$-KZ equation of level $0$: 
\begin{eqnarray}
&&F^{(l_0 ,k)}_m (\zeta_1 , \cdots , 
x^2 \zeta_j , \cdots , \zeta_{2m} )_{l_0 l_1\cdots l_{2m-1}l_0} 
= \displaystyle\sum_{l'_1 \cdots l'_{j-1} l'_{j+1} \cdots l'_{2m}} 
W'\left[ \left. \begin{array}{cc} 
l_{j} & l'_{j-1} \\ l_{j-1} & l_{j-2} \end{array} \right| 
u_j-u_{j-1}-2 \right] \nonumber 
\\[1mm]
&\times&\displaystyle\prod_{k=1}^{j-2} W'\left[ \left. \begin{array}{cc} 
l'_{k+1} & l'_{k} \\ l_{k} & l_{k-1} \end{array} \right| 
u_j-u_{k}-2 \right] W'\left[ \left. \begin{array}{cc} 
l'_{1} & l'_{2m} \\ l_{0} & l_{2m-1} \end{array} \right| 
u_j-u_{2m} \right] 
\nonumber \\[1mm] &\times & \displaystyle 
\prod_{k=j+1}^{2m} W'\left[ \left. \begin{array}{cc} 
l'_{k+1} & l'_{k} \\ l_{k} & l_{k-1} \end{array} \right| 
u_j-u_{k} \right] F^{(l'_1 , k)}_m (\zeta_1 , \cdots , 
\zeta_j , \cdots , \zeta_{2n} )_{l'_1\cdots l'_{j-1}l_jl'_{j+1}\cdots 
l'_{2m} l'_1}. \label{eq:qKZ}
\end{eqnarray}

Set 
\begin{eqnarray}
 F_{m}^{(l,k)}(\zeta )_{ll_1\cdots l_{2m-1}l}
&=&c_m \displaystyle\prod_{1\leqslant j< k \leqslant 2m} 
\zeta_j^{-\frac{r}{r-1}} g^*(z_j/z_k) \overline{F}_{m}^{(l,k)}
(\zeta)_{ll_1\cdots l_{2m-1}l}. 
\label{eq:F'-bar}
\end{eqnarray}
Here $c_m$ is a constant, and the function $g^*(z)$ is 
a scalar function defined by (\ref{eq:g*}). 

Let 
\begin{equation}
A_\pm :=\{ a|l_a =l_{a-1}\pm 1, \,\,1\leqslant a\leqslant 2m \}. 
\label{eq:df-A}
\end{equation}
Then the number of the elements of $A$ is equal to $m$ 
because $l_0 =l_{2m}$. Let us introduce the 
following meromorphic function 
\begin{equation}
\!\!\! Q'_{m}(w|\zeta )_{ll_1\cdots l_{2m-1}l}
=\displaystyle\prod_{a,b\in A_-\atop a<b} 
[v_a -v_b +1]' \!
\prod_{a\in A_-} 
\frac{[u_a -v_a -\tfrac{1}{2}+l_a ]'}{
[u_a -v_a -\tfrac{3}{2}]'} \left( 
\!\prod_{j=a+1}^{2m} \frac{[u_j -v_a -\tfrac{1}{2}]'}{
[u_j -v_a -\tfrac{3}{2}]'} \right), 
\label{eq:df-QF'} 
\end{equation}
$$
{[}u{]'}
=\displaystyle 
x^{\frac{u^2}{r-1}-u}\Theta_{x^{2(r-1)}}(x^{2u}), 
$$
where $w_a =x^{-2v_a}$ and $z_j =\zeta_j^2 =x^{-2u_j}$. 
Here we use slightly different $Q'_{m}(w|\zeta )_{ll_1\cdots l_{2m-1}l}$ 
from the one we used in \cite{SOSff}. 

The integral part $\overline{F}_{m}^{(l,k)}$ in (\ref{eq:F'-bar}) 
is given as follows: 
\begin{equation}
\overline{F}_{m}^{(l,k)} 
(\zeta)_{ll_1\cdots l_{2m-1}l}
=\prod_{a\in A_-}\oint_{C'_a} 
\dfrac{dw_a}{2\pi\sqrt{-1}w_a} 
\Psi'{}_{m}^{(i)} (w|\zeta )
Q'_{m}(w|\zeta )_{ll_1\cdots l_{2m-1}l}. 
\label{eq:F'-form}
\end{equation}
Here, $i\equiv k-l$ (mod $2$), and the kernel has the form 
\begin{equation}
\Psi'{}^{(i)}_m (w| \zeta )=
\vartheta^{(i)}_m (w | \zeta )
\prod_{a\in A_-}\prod_{j=1}^{2n} 
x^{-\frac{(v_a -u_j)^2}{2(r-1)}} 
\psi' \Bigl(\frac{w_{a}}{z_j}\Bigr) 
\prod_{1\leqslant j<k\leqslant 2n} 
x^{-\frac{(u_j -u_k)^2}{4(r-1)}},
\label{eq:df-Psi'}
\end{equation}
where
\begin{equation}
\psi' (z)=
\frac{(x^{2r+1}z;x^4,x^{2(r-1)})_{\infty}
(x^{2r+1}z^{-1};x^4,x^{2(r-1)})_{\infty}}
{(xz;x^4,x^{2(r-1)})_{\infty}(xz^{-1};x^4,x^{2(r-1)})_{\infty}}, 
\label{eq:df-psi'}
\end{equation}
\begin{equation}
\begin{array}{rcl}
\vartheta^{(i)}_m (w|\zeta )
&=&\displaystyle\left( 
(-1)^m \prod_{a\in A_-} w_a^{-1} \prod_{j=1}^{2m}\zeta_j 
\right)^i \Theta_{x^8} \left( -x^{2+4i} \prod_{a\in A_-} w_a^{-2} 
\prod_{j=1}^{2m}z_j \right) \\
&\times&\displaystyle\prod_{j=1}^{2m} 
\zeta_j^{-n(1-\frac{1}{r})-\frac{1}{2r}} \prod_{a\in A_-} x^{-mv_a} 
\prod_{a,b\in A_-\atop a<b} w_a^{-1} \Theta_{x^2} (w_a /w_b ). 
\end{array}
\label{eq:th-sol}
\end{equation}

The integrand may have poles at 
\begin{equation}
w_a =\left\{ 
\begin{array}{ll}
x^{\pm (1+4n_1+2(r-1)n_2 )}z_j & (1\leqslant j\leqslant 2m, 
n_1 , n_2\in \mathbb{Z}_{\geqslant 0}), \\
x^{3+2(r-1)n_3}z_j & (a\leqslant j\leqslant 2m, 
n_3\in \mathbb{Z}). 
\end{array} \right. 
\label{eq:pole-position'}
\end{equation}
We choose the integration contour $C'_a$ with 
respect to $w_a$ ($a\in A_-$) to be along 
a simple closed curve oriented counter-clockwise that 
encircles the points $x^{1+4n_1+2(r-1)n_2}z_j$ 
$(1\leqslant j \leqslant 2m , 
n_1 , n_2\in \mathbb{Z}_{\geqslant 0})$ 
and $x^{3+2(r-1)n_3}z_j$ $(a\leqslant j \leqslant 2m , 
n_3\in \mathbb{Z}_{>0})$, 
but not $x^{-1-4n_1-2(r-1)n_2} z_j$ 
$(1\leqslant j \leqslant 2m, 
n_1 , n_2\in \mathbb{Z}_{\geqslant 0})$ 
nor $x^{3-2(r-1)n_3}z_j$ $(a\leqslant j \leqslant 2m, 
n_3\in \mathbb{Z}_{\geqslant 0})$. Thus, the contour $C'_a$ actually 
depends on the variables $z_j$, and therefore 
strictly, it should be written $C'_a (z)$. 
The LHS of (\ref{eq:F'cyc-comp}) represents the analytic 
continuation with respect to $\zeta_{2m}$. 

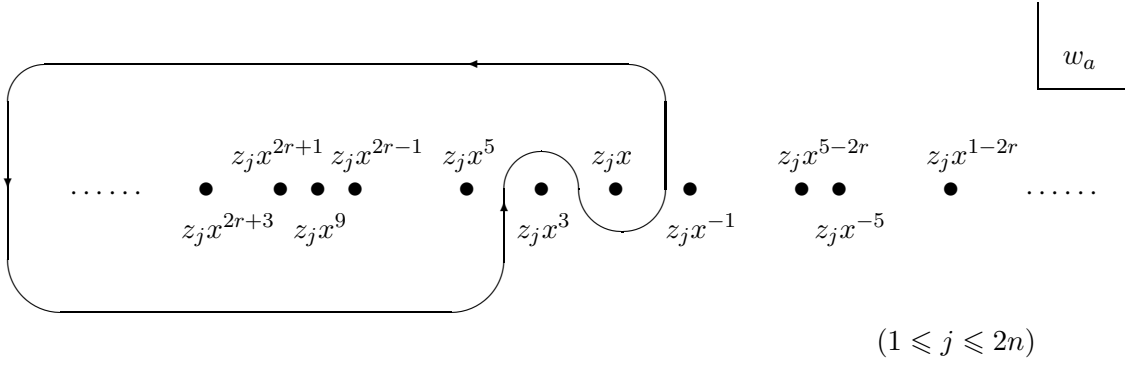
\unitlength 1.65mm
\begin{picture}(140,35)
\put(-27,-13){
\put(45,25){\oval(40,20)[b]}
\put(28,35){\line(1,0){44}}
\put(28,32){\oval(6,6)[tl]}
\put(25,32){\vector(0,-1){7}}
\put(62,25){\circle*{1}}
\put(53,25){\circle*{1}}
\put(50,25){\circle*{1}}
\put(47,25){\circle*{1}}
\put(41,25){\circle*{1}}
\put(30,24.1){$\cdots\cdots$}
\put(68,25){\circle*{1}}
\put(74,25){\circle*{1}}
\put(80,25){\circle*{1}}
\put(89,25){\circle*{1}}
\put(92,25){\circle*{1}}
\put(101,25){\circle*{1}}
\put(107,24.1){$\cdots\cdots$}
\put(60,27){$z_j x^{5}$}
\put(48,21){$z_j x^9$}
\put(51,27){$z_j x^{2r-1}$}
\put(43,27){$z_j x^{2r+1}$}
\put(39,21){$z_j x^{2r+3}$}
\put(66,21){$z_j x^{3}$}
\put(72,27){$z_j x$}
\put(78,21){$z_j x^{-1}$}
\put(87,27){$z_j x^{5-2r}$}
\put(99,27){$z_j x^{1-2r}$}
\put(90,21){$z_j x^{-5}$}
\put(65,19){\vector(0,1){5}}
\put(68,25){\oval(6,6)[t]}
\put(74.5,25){\oval(7,7)[b]}
\put(78,25){\line(0,1){7}}
\put(75,32){\oval(6,6)[tr]}
\put(75,35){\vector(-1,0){13}}
\put(95,12){\normalsize($1\leqslant j\leqslant 2n$)}
\put(110,35){$w_a$}
\put(108,33){\line(0,1){7}}
\put(108,33){\line(1,0){7}}
}
\end{picture}

~

Then $F_{m}^{(l,k)}(\zeta )_{ll_1\cdots l_{2m-1}l}$ 
satisfies level $0$ $q$-KZ equations \cite{SOSff}, and therefore 
it can be identified a form factor in the eight-vertex SOS model. 

\section{Form factors in the eight-vertex model}

Let us introduce $F_m^{(i)} (\zeta )$, 
the form factors in the eight-vertex model 
through the vertex-face transformation as follows: 
\begin{eqnarray}
F_{m}^{(l_0 ,k)}(\zeta )_{l_0l_1\cdots l_{2m-1}l_{2m}}
=\displaystyle\sum_{\mu_1 , \cdots , \mu_{2m}} 
F_m^{(i)} (\zeta )_{
\mu_1 \cdots \mu_{2m}} t'{}^{l_1}_{l_0} (u_1 -u_0 )^{\mu_1} 
\cdots t'{}^{l_{2m}}_{l_{2m-1}} (u_{2m} -u_0 )^{\mu_{2m}}. 
\label{eq:F-vf} 
\end{eqnarray}
Here $i\equiv k-l_0$ (mod $2$), and  
\begin{equation}
t'{}^{k}_{k'} (u):=t^{k}_{k'} (u; \epsilon , r-1). 
\label{eq:t'}
\end{equation}
Let us remind (\ref{eq:t'*}) and let us introduce 
\begin{equation}
\tilde{t}'{}^{*k}_{k'} (u):
=\tilde{t}^{*k}_{k'} (u; \epsilon , r-1). 
\label{eq:t'*t}
\end{equation}
Then the following inversion relations hold: 
\begin{equation}
\hspace{-1.0mm}\sum_{\varepsilon =\pm} 
t'{}^{*k'}_{k} (u)_\varepsilon 
t'{}^{k}_{k''} (u)^\varepsilon =\delta_{k''}^{k'}, ~~~~
\sum_{k'=k\pm 1} t'{}^{k}_{k'} (u)^\varepsilon 
t'{}^{*k'}_{k} (u)_{\varepsilon'} =
\delta^\varepsilon_{\varepsilon'}, \label{eq:dual-t*}
\end{equation}
\begin{equation}
\hspace{-1.0mm}\sum_{\varepsilon =\pm} 
\tilde{t}'{}_{k'}^{*k} (u)_\varepsilon 
t'{}^{k''}_{k} (u)^\varepsilon =\delta^{k''}_{k'}, ~~~~
\sum_{k'=k\pm 1} t'{}_{k}^{k'} (u)^\varepsilon 
\tilde{t}'{}_{k'}^{*k} (u)_{\varepsilon'} =
\delta^\varepsilon_{\varepsilon'}. \label{eq:dual-t'*}
\end{equation}
It follows from (\ref{eq:dual-t*}) and (\ref{eq:dual-t'*}) 
that the relation (\ref{eq:F-vf}) is equivalent to 
\begin{equation}
\begin{array}{rcl}
F_m^{(i)} (\zeta )&=&\displaystyle\sum_{
l_0 , \cdots , l_{2m-1}} F_{m}^{(l_0 ,k)}(\zeta )_{
l_0l_1\cdots l_{2m-1}l_{2m}} t'{}^{*l_0}_{l_1} (u_1 -u_0 ) 
\otimes \cdots \otimes t'{}^{*l_{2m-1}}_{l_{2m}} 
(u_{2m} -u_0 ) \\[5mm]
&=&\displaystyle\sum_{
l_1 , \cdots , l_{2m}} F_{m}^{(l_0 ,k)}(\zeta )_{
l_0l_1\cdots l_{2m-1}l_{2m}} 
\displaystyle \tilde{t}'{}^{*l_0}_{l_1} (u_1 -u_0 ) 
\otimes \cdots \otimes \tilde{t}'{}^{*l_{2m-1}}_{l_{2m}} 
(u_{2m} -u_0 ). \end{array}
\label{eq:XYZff}
\end{equation}
Thus, the $S$-matrix symmetry (\ref{eq:S-symm}) 
for $F_m^{(i)} (\zeta )$ follows from 
the $W'$-symmetry (\ref{eq:W'-symm-comp}) 
for $F_{m}^{(l_0 ,k)}(\zeta )$. 

It is evident from (\ref{eq:dual-t*}) and (\ref{eq:dual-t'*}) 
that one of the sufficient conditions of (\ref{eq:cyc-fi}), 
the cyclicity for $F_m^{(i)} (\zeta )$, is as follows: 
\begin{eqnarray}
&&\displaystyle\sum_{l_{2m}=\atop l_{2m-1}\pm 1} \tilde{t}'{}^{
*l_{2m-1}}_{l_{2m}}(u_{2m}-u_0 +2)_\mu 
F_{m}^{(l_0 ,k)}(\zeta', x^{-2}\zeta_{2m})_{
l_0\cdots l_{2m-1}l_{2m}} \nonumber \\[1mm]
&=&\displaystyle\sum_{l'=l_0 \pm 1} t'{}^{*l'}_{l_0}
(u_{2m}-u_0 )_\mu  F_{m}^{(l',k)}(\zeta_{2m}, \zeta' )_{
l'l_0\cdots l_{2m-1}}. \label{eq:cyc-gen}
\end{eqnarray}
The strategy is as follows. We have an expression for only 
the case $l_{2m}=l_0$. Thus, first let $l_{2m-1}=l_0 \pm 1$ 
and solve (\ref{eq:cyc-gen}). Then we will 
obtain formulae for $l_{2m}=l_0 \pm 2$. 
Next let $l_{2m-1}=l_0 \pm 3$ and solve (\ref{eq:cyc-gen}). 
Then we will obtain formulae for $l_{2m}=l_0 \pm 4$. 
Repeating this procedure, we will obtain the general formulae 
for $l_{2m}\equiv l_0$ (mod $2$). 

For generic $r$, not (\ref{eq:cyc-gen}) 
but (\ref{eq:cyc-rless}) does holds: 
\begin{eqnarray}
&&\displaystyle\sum_{l_{2m}=\atop l_{2m-1}\pm 1} \tilde{t}'{}^{
*l_{2m-1}}_{l_{2m}}(u_{2m}-u_0 )_\mu 
F_{m}^{(l_0 ,k)}(\zeta', x^{-2}\zeta_{2m})_{
l_0l_1\cdots l_{2m-1}l_{2m}} \nonumber \\[1mm]
&=&\displaystyle\sum_{l'=l_0 \pm 1} t'{}^{*l'}_{l_0}
(u_{2m}-u_0 )_\mu  F_{m}^{(l',k)}(\zeta_{2m}, \zeta' )_{
l'l_0l_1\cdots l_{2m-1}}. \label{eq:cyc-rless}
\end{eqnarray}
Here, for $l_{2m}=l+2s\geqslant l$, let 
$$
A'_- =A_- \sqcup \{ -1, \cdots , -s\}, 
$$
and $l_{-i}=l+2(i-1)$ for $1\leqslant i\leqslant s$. Then 
the meromorphic function $Q'_{m}(w|\zeta )_{ll_1\cdots l_{2m-1}l+2s}$ 
is defined as follows \cite{Qbk}: 
\begin{equation}
\begin{array}{cl}
&Q'_{m}(w|\zeta )_{ll_1\cdots l_{2m-1}l+2s}
=\displaystyle\prod_{a\in A_-} 
\frac{[u_a -v_a -\tfrac{1}{2}+l_a ]'}{
[u_a -v_a -\tfrac{3}{2}]'} \left( 
\prod_{j=a+1}^{2m} \frac{[u_j -v_a -\tfrac{1}{2}]'}{
[u_j -v_a -\tfrac{3}{2}]'} \right) \\[5mm]
\times&\displaystyle\prod_{a,b\in A'_-\atop a<b} 
[v_a -v_b +1]' 
\displaystyle\prod_{a'=-1}^{-s} 
\frac{[u_0 -v_{a'}-\tfrac{1}{2}+l_{a'}]'}{
[u_0 -v_{a'}-\tfrac{3}{2}]'} \left( 
\prod_{j=1}^{2m} \frac{[u_j -v_{a'}-\tfrac{1}{2}]'}{
[u_j -v_{a'}-\tfrac{3}{2}]'} \right). 
\end{array}
\label{eq:df-QF'+}
\end{equation}
The meromorphic function 
$Q'_{m}(w|\zeta )_{ll_1\cdots l_{2m-1}l-2s}$ for 
$l_{2m}=l-2s\leqslant l$ can be defined similarly, see \cite{Qbk}. 
In order to derive (\ref{eq:cyc-rless}) we use the relation 
(\ref{eq:C-df}) with $r$ replaced by $r-1$, and the 
addition theorems 
$$
\begin{array}{cl}
&\theta_i \left( \tfrac{l+2s-1-(u-u_0 )}{2(r-1)}; 
\tfrac{\pi\sqrt{-1}}{2\epsilon (r-1)} \right) 
\frac{[u-v-\tfrac{1}{2}+l+2(s-1)]'}{
[u-v-\tfrac{3}{2}]'} 
-\theta_i \left( \tfrac{l+2s-1+(u-u_0 )}{2(r-1)}; 
\tfrac{\pi\sqrt{-1}}{2\epsilon (r-1)} \right) \\[3mm] 
\times&
\frac{[u_0 -v-\tfrac{1}{2}+l+2(s-1)]'}{
[u_0 -v-\tfrac{3}{2}]'} = \frac{[u_0 -u]'[l+2s-1]'}{
[u_0 -v-\tfrac{3}{2}]'[u-v-\tfrac{3}{2}]'}
\theta_i \left( \tfrac{l+u+u_0 -2v+2s-4}{2(r-1)}; 
\tfrac{\pi\sqrt{-1}}{2\epsilon (r-1)} \right), \\[5mm]
&\theta_i \left( \tfrac{l-(u-u_0 +1)}{2(r-1)}; 
\tfrac{\pi\sqrt{-1}}{2\epsilon (r-1)} \right) 
\frac{[u-v-\tfrac{1}{2}+l]'}{
[u-v-\tfrac{3}{2}]'} -
\theta_i \left( \tfrac{l+(u-u_0 +1)}{2(r-1)}; 
\tfrac{\pi\sqrt{-1}}{2\epsilon (r-1)} \right) \\[3mm]
\times&\frac{[u_0 -v-\tfrac{3}{2}+l]'}{
[u_0 -v-\tfrac{3}{2}]'}
\frac{[u-v-\tfrac{1}{2}]'}{[u-v-\tfrac{3}{2}]'} 
=\frac{[u_0 -u-1]'[l]'}{
[u_0 -v-\tfrac{3}{2}]'[u-v-\tfrac{3}{2}]'}
\theta_i \left( \tfrac{l+u+u_0 -2v-2}{2(r-1)}; 
\tfrac{\pi\sqrt{-1}}{2\epsilon (r-1)} \right), \end{array}
$$
where $i=3,4$. 

When $r=r_N =1+1/N$ ($N=1,2,3,\cdots$), the eight-vertex model 
is called reflectionless. At $r=r_N$, 
$$
\tilde{t}'{}^{
*l'}_{l}(u)=\tilde{t}'{}^{*l'}_{l}(u+2)
$$
holds. Thus, (\ref{eq:cyc-rless}) implies (\ref{eq:cyc-gen}) 
at $r=r_N$. 
Furthermore, the sum with respect to $l_j$ can be carried out 
when $r=r_N$, by rewriting $F_{m}^{(l,k)}$ as $2m$-fold 
integral form: 
\begin{eqnarray}
\overline{F}_{m}^{(i)} 
(\zeta)_{\mu_1\cdots \mu_{2m}} 
&=&\displaystyle\prod_{a=1}^{2m}\oint_{C'} 
\dfrac{dw_a}{2\pi\sqrt{-1}w_a} 
\Psi{}_{m}^{(i)} (w|\zeta )
Q^{(i)}_{m}(w|\zeta )_{\mu_1\cdots \mu_{2m}}. 
\label{eq:F-form} 
\end{eqnarray}

Here, 
\begin{eqnarray}
Q^{(i)}_{m}(w|\zeta )_{\mu_1\cdots \mu_{2m}} 
&=&\displaystyle\prod_{a=1}^{2m} \dfrac{1}{[u_0 -v_a -\frac{3}{2}]'
[u_a -v_a -\frac{3}{2}]'} 
\displaystyle\left( \prod_{j=a+1}^{2m}\frac{[u_j -v_a -\frac{1}{2}]'}
{[u_j -v_a -\frac{3}{2}]'}\right) \nonumber \\
&\times&\displaystyle\prod_{j=1}^{2m} \theta_{\overline{\mu_j}} 
\left( \tfrac{i+u+u_0 -2v}{2(r-1)}; 
\tfrac{\pi\sqrt{-1}}{2\epsilon (r-1)} \right) 
\prod_{a<b\atop a,b=1}^{2m} [v_a -v_b +1]', \label{eq:sumpath} 
\end{eqnarray}
and $\overline{+}=3$, $\overline{-}=4$. 

The resulting formulae suggest us that the free 
field representation of the type II vertex operators\footnote{
Concerning the terminology type I and II, see e.g., \cite{JMbk}. } 
are as follows: 
\begin{equation}
\begin{array}{rcl}
\Psi^{*(1-i,i)}_\mu (\zeta )&=&
\displaystyle\oint_{C'} 
\dfrac{dw}{2\pi\sqrt{-1}w} \psi^*(\zeta )B(w) 
\dfrac{\theta_{\bar{\mu}}\left( \tfrac{i+u+u_0 -2v}{2(r-1)}; 
\tfrac{\pi\sqrt{-1}}{2\epsilon (r-1)} \right)}{
[u_0-v-\frac{3}{2}]'[u-v-\frac{3}{2}]'}. \end{array}
\end{equation}
where 
$$
\psi^*(\zeta )=\zeta^{\frac{r}{2(r-1)}}:\exp \left( 
\sqrt{\dfrac{r}{2(r-1)}} (\sqrt{-1}Q+P\log z)+
\sum_{m\neq 0} \frac{\alpha_m}{m}z^{-m} \right), 
$$
$$
B(w)=w^{\frac{r}{(r-1)}}:\exp \left( 
-\sqrt{\dfrac{2r}{(r-1)}} (\sqrt{-1}Q+P\log w)-
\sum_{m\neq 0} \frac{\alpha_m}{m}\frac{[2m]_x}{[m]_x}z^{-m} \right). 
$$
Here we use the bosonic oscillators with the following 
commutation relations: 
\begin{equation}
\begin{array}{rcl}
[\alpha_m , \alpha_n ]&=&m\dfrac{[m]_x [rm]_x}{[2m]_x [(r-1)m]_x} 
\delta_{m+n,0}, ~~~~ [m]_x :=\dfrac{x^m -x^{-m}}{x-x^{-1}}, \\[3mm]
{[} Q,P {]}&=&\sqrt{-1}. \end{array}
\end{equation}

\section{Summary and discussion}

In this paper, we tried to construct the form factors in 
the eight-vertex model as solutions to level $0$ 
$q$-KZ equation, or Smirnov's axioms. The $q$-KZ equation 
was reduced to (\ref{eq:cyc-gen}). 
Up to now, eq. (\ref{eq:cyc-gen}) has been solved only at 
reflectionless points $r=1+1/N$ ($N=1,2,3,\cdots$). 
On these points, we further succeeded to construct the 
free field representation of the type II vertex operators. 

Let us list a few open problems. 

\noindent 1) Obtain the type I vertex operators 
at reflectionless points, which should commute the type II ones 
with some scalars, and which themselves should 
satisfy appropriate commutation relations. 

\noindent 2) Solve (\ref{eq:cyc-gen}) for generic $r>1$. 

\noindent 3) Find the link with Shiraishi's work, 
in which the type I and II vertex operators can be 
constructed from the representations of the 
deformed $W(D_{N+1})$ or $W(B_l^{(m)}\otimes B_m )$ algebra. 

Shiraishi's bosonization is phenomenological in the sense 
that the relation between the eight-vertex model and 
the deformed $W$ algebra is unclear, at least up to now. 
In a joint work with M. Lashkevich, we study to show 
that the form factors at reflectionless points can be 
obtained without integrals on the basis of vertex-face 
transformation method. Throughout this study, we wish to 
give theoretical account of Shiraishi's scheme. 

\section*{Acknowledgements}
The author would like to thank M. Jimbo, H. Konno, 
M. Lashkevich, A. Nakayashiki, Ya. Pugai and J. Shiraishi 
for useful discussion. 
This work was supported in part by a Grant-in-Aid for 
Scientific Research from JSPS, Japan Society for 
the Promotion of Science (No. 15540218).


\begin{thebibliography}{99}
\bibitem{Smbk}Smirnov, F. A. {\it 
Form factors in completely integrable models of
quantum field theory}; Advanced Series in Mathematical 
Physics Vol {\bf 14}; World Scientific: Singapore, 1992.
\bibitem{FR}Frenkel I. B. ; Reshetikhin, N. Y. 
Quantum affine algebras and holonomic difference equations. 
{\it Commun. Math. Phys.} 1992, {\bf 146}, 1--60.
\bibitem{LaP}Lashkevich, M. ; Pugai, Ya. 
Free field construction for correlation 
functions of the eight vertex model. 
{\it Nucl. Phys.} 1998, {\bf B516}, 623--651. 
\bibitem{LaP2}Lashkevich, M. ; Pugai, Ya. 
Nearest neighbor two-point correlation function of the $Z$-invariant 
eight vertex model. {\it JETP Lett.} 1998, {\bf 68}, 257--262. 
\bibitem{3-bu/2}Baxter, R. J. Eight-vertex model in lattice 
statistics and one-dimensional anisotropic 
Heisenberg chain. I. Some fundamental eigenvectors. 
{\it Ann. Phys.} (NY) 1973, {\bf 76}, 1--24; II. 
Equivalence to a generalized ice-type lattice model. 
{\it ibid.}, 25--47; 
III. Eigenvectors of the transfer matrix and 
Hamiltonian. {\it ibid.}, 48--71. 
\bibitem{LuP}Lukyanov, S. ; Pugai, Ya. 
Multi-point local height probabilities in the 
integrable RSOS model. 
{\it Nucl. Phys.} 1996, {\bf B473}[FS], 631--658. 
\bibitem{qXYZ}Quano, Y-H. Bootstrap equations and 
correlation functions for the Heisenberg XYZ antiferromagnet. 
{\it J. Phys. A: Math. Gen.} 2002, {\bf 35}, 9549--9572. 
\bibitem{Shi}Shiraishi, J. (2003) Free field constructions for 
the elliptic algebra ${\cal A}_{q,p}(\widehat{\mathfrak{sl}_2})$ 
and Baxter's eight-vertex model. math.QA/0302097. 
\bibitem{La}Lashkevich, M. Free field construction 
for the eight-vertex model: representation for form factors. 
{\it Nucl. Phys.} 2002, {\bf B621}, 587--621. 
\bibitem{SOSff}Quano, Y-H. Quantum 
Knizhnik--Zamolodchikov equations of level $0$ and 
form factors in SOS model. 
{\it Prog. Theo. Phys.} 2004, {\bf 111}, 361--370. 
\bibitem{Zinv}Baxter, R. J. Solvable eight-vertex model 
on an arbitrary planar lattice. {\it Phil. Trans. Roy. Soc.} 
(London) 1978, {\bf 289A}, 315--346. 
\bibitem{ESM}Baxter, R. J. {\it Exactly Solved Models 
in Statistical Mechanics}; Academic Press: London, 1982.
\bibitem{Qbk}Quano, Y-H. 
Difference equations for correlation functions and 
form factors of the eight-vertex/XYZ model, to appear in 
{\it Progress in Ferromagnetism Research}, 
Nova Science Publ: NY. 
\bibitem{JMbk}Jimbo, M. ; Miwa, T. 
{\it Algebraic analysis of solvable lattice models}; 
CBMS Regional Conferences Series in Mathematics Vol {\bf 85}; 
AMS: Providence, RI, 1994. 
\end{thebibliography}
\end{document}